\documentclass[a4paper]{article}

\usepackage{amsmath}
\usepackage{graphicx}
\usepackage{graphicx}

\makeatletter

\@addtoreset{equation}{section}
\makeatother

\newcommand{\Dbar}{\overline{\mathcal{D}}}
\newcommand{\Fbar}{\overline{\mathcal{F}}}
\newcommand{\Ubar}{\,\overline{\mathcal{U}}}

\newcommand{\tr}{\mathop{\mathrm{tr}}\nolimits}

\begin{document}


\begin{center}
\renewcommand{\thefootnote}{\fnsymbol{footnote}}
\newcommand{\inst}[1]{\mbox{$^{\text{\textnormal{#1}}}$}}
{\LARGE
Lattice formulation of two-dimensional $\mathcal{N}=(2,2)$
super Yang-Mills with $SU(N)$ gauge group
}\\[8ex]
%
{\large
Issaku Kanamori\footnote{\texttt{issaku.kanamori@physik.uni-regensburg.de}}
}\\
%
{\large\itshape
Institut f\"ur Theoretische Physik, Universit\"at Regensburg,\\
D-93040 Regensburg, Germany
}
\end{center}
\bigskip\bigskip
\setcounter{footnote}{0}


\begin{abstract}
 We propose a lattice model for two-dimensional 
$SU(N)$ $\mathcal{N}=(2,2)$ super Yang-Mills model.
We start from the CKKU model for this system, which is valid only for 
$U(N)$ gauge group.  We give a reduction of $U(1)$ part keeping
a part of supersymmetry.  In order to suppress artifact vacua, we use
an admissibility condition.
\end{abstract}

\newpage

\section{Introduction}

Supersymmetric gauge theories play an important role 
in both phenomenological and purely theoretical aspects.
It is very natural to try to find a way to define supersymmetric
theories nonperturbatively: a lattice regularization is a nice candidate.
There have been proposed several approaches to the lattice regularization
of supersymmetric Yang-Mills theories in the past decade
(For reviews see \cite{Catterall:2009it, Joseph:2011xy}, for example).
Most of them possess at least one exact supertransformation,
which has an interpretation of a scalar supercharge in terms of
topological twist. The supersymmetry (SUSY) algebra
contains \emph{infinitesimal} translations but 
the lattice allows only finite translations: 
the SUSY algebra needs to be represented by the finite translation.
The (full) SUSY is broken at the finite lattice spacing,
without introducing non-standard properties such as non-locality~\cite{
Kato:2008sp,Bergner:2009vg}.\footnote{
In low dimensional non-gauge systems, 
regularizations through momentum space can give 
exact full SUSY at finite cutoff, of the price of non-locality
but becomes local in the infinite cutoff limit~\cite{Kadoh:2009sp}.  
Different models with non-locality are 
found in~\cite{D'Adda:2010pg,D'Adda:2011jw}. 
}

In two dimensions the above mentioned exact scalar symmetry is strong
enough to guarantee a restoration of full supersymmetry without fine
tuning~\cite{Kaplan:2002wv}, which was explicitly checked 
in Monte Carlo simulations~\cite{Kanamori:2008bk} for 
a model by Sugino~\cite{Sugino:2004qd}.  
In one dimension, 
a non-lattice approach without any exact supersymmetry at
finite cutoff provides fine tuning free regularization~\cite{Hanada:2007ti}.
Other fine tuning free lattice/non-lattice models for gauge theories
are found 
in~\cite{Hanada:2010kt,Hanada:2010gs,Hanada:2011qx,Ishii:2008ib,Ishiki:2011ct}.

It is interesting to note that
some of the known lattice formulations treat fermions as link variables.
Since the gauge fields are treated as link variables on the lattice, 
it is quite natural 
to introduce fermions on links as superpartner of bosonic link variables.
Here we focus on a 2-dimensional $\mathcal{N}=(2,2)$ system,
which is the well studied system on the lattice.
A Model proposed in \cite{Cohen:2003xe} (CKKU model)
was derived from a matrix model by using
orbifolding, which naturally gives fermionic link variables.
A geometrical approach \cite{hep-lat/0410052} also uses link fermions.
In \cite{D'Adda:2005zk}, the present author together with his collaborators
tried to introduce supercharges on links as well (link approach).
The link approach originally intended to keep the full exact
supersymmetry on the lattice at finite lattice spacing,
but was turned out to be equivalent to the CKKU model~\cite{Damgaard:2007eh}.
On the other hand, a model proposed by Sugino (Sugino model), 
which also keeps
the exact scalar supercharge, 
treat the fermion as site variables~\cite{Sugino:2003yb, 
Sugino:2004qd, Sugino:2006uf}.
Both the CKKU model and the Sugino model describe the same target system
in the continuum limit.  In fact, both models give the same numerical
results~\cite{Hanada:2010qg}. 
See \cite{hep-lat/0507019, Sugino:2006uf,arXiv:0705.3831, Kadoh:2009yf} 
for other
approaches to this system and relation among the formulations,
and \cite{Catterall:2011cea, Catterall:2011aa} for
recent numerical studies.
Note that this system is a 2-dimensional cousin of 4-dimensional
$\mathcal{N}=4$ system in terms of Dirac-K\"ahler twist
\cite{hep-th/9905222, hep-th/0310242}.

There are two types of 
topological twist in 2-dimensional $\mathcal{N}=(2,2)$ systems
and CKKU and Sugino models use different ones.
They are called A-model twist and B-model twist.
A-model twist combines the spacetime rotation and the internal $U(1)_V$
rotation.  
B-model uses the internal $U(1)_A$ instead of $U(1)_V$.
CKKU model uses B-model twist and Sugino model
uses A-model twist. 
We list further differences between the two formulations 
in Table~\ref{tab:difference1}.
\begin{table}[htbp]
 \hfil
 \begin{tabular}{c|c|c}
  & Sugino& CKKU \\
 \hline
 twist & A-model & B-model \\
 fermion & site & link \\
 gauge group & $U(N)$ or $SU(N)$ & $U(N)$ only \\
 admissibility & needed & no need
 \end{tabular}
\caption{Comparison of the Sugino model and the CKKU model.}
\label{tab:difference1}
\end{table}

In pure $U(N)$ super Yang-Mills theory without any matter multiplets,
the $U(1)$ part of the gauge group is decoupled from the other part
of the dynamics.  However, in the CKKU model,
due to the lattice artifacts
the decoupling is not complete at finite lattice spacing.  
The CKKU model allows only $U(N)$ gauge group by construction
and in fact a naive reduction to $SU(N)$ by hand does not work 
(see Sec.~\ref{sec:sun}).
The coupling with the $U(1)$ part may cause
a fake sign problem as well~\cite{Hanada:2010qg}.
Another, and more crucial problem of the $U(1)$ part
of CKKU model is a stability of the $U(1)$ part of the scalar field.
Because its expectation value gives the lattice spacing,
the stability is quite important.
An early analysis on this issue is found in~\cite{Onogi:2005cz}.
In Monte Carlo simulations, an ad hoc treatment might be needed
to stabilize it.
One practical way is to introduce a mass term specific to the $U(1)$ 
part~\cite{Hanada:2010qg}.
On the other hand, 
the Sugino model with $SU(N)$ gauge group is free from these problems.
The cost we have to pay for the Sugino model is 
an admissibility condition, which is needed to
suppress unphysical artifact vacua.  The action is thus more complicated
than that of CKKU model and
an implementation of the model for numerical simulation
becomes more complicated as well.

Motivated by the simpler implementation but the rather complicated
treatment of $U(1)$ part of the CKKU model,
in this paper, we propose an $SU(N)$ version of CKKU model.
As we will describe, the obtained model is rather close to
the Sugino model.
Because link fermions require $U(N)$ gauge group,
we need to use site fermions.  In order to suppress artifact vacua we
need the admissibility conditions as well.  
Unfortunately, because of the admissibility
condition, the action becomes complicated.

In the next section we give a brief review of the CKKU model which uses
$U(N)$ gauge group.
Then we reduce the gauge group to $SU(N)$ in Sec.~\ref{sec:sun}.
The (classical) vacuum structure is analyzed in Sec.~\ref{sec:vacua},
which is needed for the admissibility condition.
Sec.~\ref{sec:conclusions} contains conclusions and discussions.

\section{A Brief Review of the $U(N)$ Model}
\label{sec:un}

In this section we give a brief review of the lattice model
introduced in \cite{Cohen:2003xe} and settle the notations.
The target system is 2-dimensional $\mathcal{N}=(2,2)$ 
super Yang-Mills theory.
We do not follow the original derivation with orbifolding
and deconstruction but put emphasis on the nilpotent $Q$-symmetry.

\begin{figure}
 \hfil\includegraphics{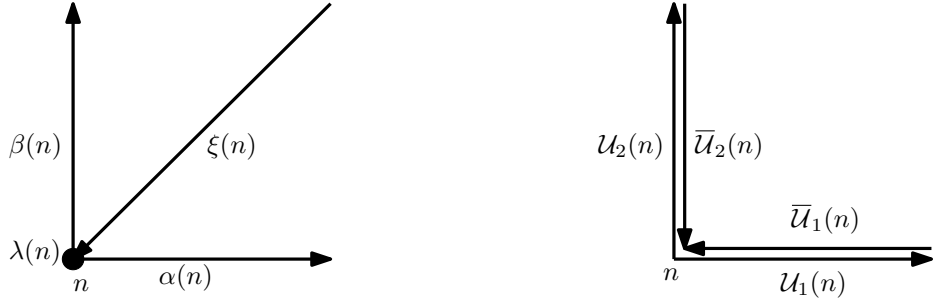}
   \caption{Fields on the lattice for the CKKU model: fermions (left) and bosons (right).}
 \label{fig:ckku_fields}
\end{figure}

We denote complex boson fields, which are made of gauge fields
and scalar fields, as $\mathcal{U}_\mu$ and 
$\Ubar_\mu (=\mathcal{U}_\mu^\dagger)$.  We set all fields dimensionless.%
\footnote{Relations to the original notation in \cite{Cohen:2003xe}
are the following, where $a$ is the lattice spacing:
\begin{align*}
 \text{bosons}&\qquad
 \mathcal{U}_\mu 
   = \sqrt{2}\,a( x_{\rm CKKU}, y_{\rm CKKU} )\\
 \text{aux. field} &\qquad
  d  = a^2 d_{\rm CKKU} \\
 \text{fermions}&\qquad
  \alpha,\beta,\lambda,\xi 
   =a^\frac{3}{2}
    (\alpha_{\rm CKKU},
     \beta_{\rm CKKU},
     \lambda_{\rm CKKU},
     \xi_{\rm CKKU})\\
 \text{scalar super trans.} &\qquad
  Q  = a^\frac{1}{2} Q_{\rm CKKU}
\end{align*}
}
They live on links and their gauge transformations are
\begin{equation}
 \mathcal{U}_\mu(n) \rightarrow G(n)\mathcal{U}_\mu(n)G^{-1}(n+\hat\mu),
 \label{eq:link-gauge-trans}
\end{equation}
where $G(n)$ is a group element of the gauge group,
$n$ is a lattice site,
and $\hat\mu$ is a unit vector in $\mu$-th direction.
A bosonic auxiliary field $d$ is assigned to sites so transforms as site
variable:
\begin{equation}
 d(n) \rightarrow G(n)d(n)G^{-1}(n).
\end{equation}
Fermions $\alpha,\ \beta,\ \xi$ are also assigned to links and $\lambda$
is to sites, so the gauge transformation reads:
\begin{align}
 \alpha(n) &\rightarrow G(n)\alpha(n)G^{-1}(n+\hat1), \\
 \beta(n)  &\rightarrow G(n)\beta(n)G^{-1}(n+\hat2), \\
 \xi(n)    &\rightarrow G(n+\hat1+\hat2)\xi(n)G^{-1}(n), \\
 \lambda(n)    &\rightarrow G(n)\lambda(n)G^{-1}(n).
\end{align}
See Fig.~\ref{fig:ckku_fields}.
In terms of topological twist they are in the twisted basis:
$\alpha$ and $\beta$ make a 2-dimensional vector, 
$\xi$ is an anti-symmetric tensor (i.e.,
a pseudo scalar in two dimensions), and $\lambda$ is a scalar.

\begin{figure}
 \hfil\includegraphics{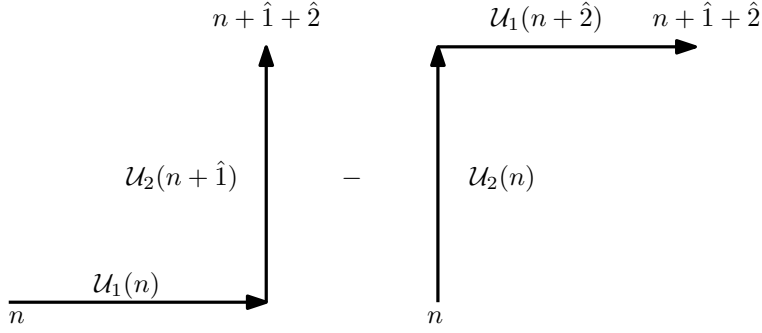}
 
 \caption{A shifted commutator 
$[\mathcal{U}_1, \mathcal{U}_2]'(n,n+\hat1+\hat2)
 =\mathcal{U}_1(n)\, \mathcal{U}_2(n+\hat1) 
 - \mathcal{U}_2(n)\, \mathcal{U}_1(n+\hat2)$.}
 \label{fig:shifted-commutator}
 
\end{figure}

They transform under a fermionic transformation $Q$,
which is a scalar part of the twisted supersymmetry,
in the following way:
\begin{align}
 Q \mathcal{U}_1(n) &= 2\alpha(n), \qquad
 Q \alpha(n)=0, \label{eq:QU1}\\*
 Q \mathcal{U}_2(n) &= 2\beta(n), \qquad
 Q \beta(n) =0, \label{eq:QU2}\\
 Q \Ubar_1(n) &= Q\Ubar_2(n) =0, \\
 Q \lambda(n) 
  &= -\frac{1}{2}\Bigl(
      \Ubar_1(n-\hat1)\mathcal{U}_1(n-\hat1) 
      - \mathcal{U}_1(n) \Ubar_1(n) \nonumber\\
  &\qquad
      + \Ubar_2(n-\hat2)\mathcal{U}_2(n-\hat2) 
      - \mathcal{U}_2(n) \Ubar_2(n)
     \Bigr)
      -id(n) \nonumber\\*
  &= -\frac{1}{2}\left(
      [\Ubar_1, \mathcal{U}_1]'(n,n) 
       + [\Ubar_2, \mathcal{U}_2]'(n,n) \right)
      -id(n), \\
  Q d(n)
   &= i \left(\Ubar_1(n-\hat1)\alpha(n-\hat1) 
              -\alpha(n)\Ubar_1(n) \right)
      +i \left(\Ubar_2(n-\hat2)\beta(n-\hat2) 
              -\beta(n)\Ubar_2(n) \right) \nonumber\\*
   &= i\left([\Ubar_1, \alpha]'(n,n)
            + [\Ubar_2, \beta]'(n,n) \right), \\
  Q\xi(n)
   &= \Ubar_1(n+\hat2)\Ubar_2(n)
       -\Ubar_2(n+\hat1)\Ubar_1(n) \nonumber \\*
   &= [\Ubar_1, \Ubar_2]'(n+\hat1+\hat2,n). \label{eq:Qxi}
\end{align}
Here we have introduced a shifted commutator 
\begin{equation}
 [A,B]'(n,m) \equiv A(n,n+a_A) B(n+a_A, m) - B(n,n+a_B) A(n+a_B,m),
\end{equation}
where $n+a_A+a_B=m$ and 
the locations of $A$ and $B$ are shifted to keep the gauge
covariance (see Fig.~\ref{fig:shifted-commutator}).
Their gauge transformations are
\begin{align}
 A(n,n_A)& \rightarrow G(n)A(n,n_A)G^{-1}(n_A),
 &   B(n,n_B)& \rightarrow G(n)B(n,n_B)G^{-1}(n_B),
\end{align}
respectively.
The argument $(n,m)$ refers to the starting and end point of the link 
and the shifted commutator transforms as
\begin{equation}
 [A,B]'(n,m)\rightarrow G(n)[A,B]'(n,m) G^{-1}(m).
\end{equation}
It is easy to check that the above $Q$-transformation is nilpotent ($Q^2=0$).

The action is given in a $Q$-exact form:
\begin{equation}
 S_{U(N)}
 = Q\Lambda_{U(N)} \label{eq:un-action}
\end{equation}
with preaction
\begin{align}
\Lambda_{U(N)}
 &= \kappa \sum_n \tr\Bigl[
    -\frac{1}{4}\lambda(n)\left\{
      [\Ubar_1,\mathcal{U}_1]'(n,n)
     + [\Ubar_2, \mathcal{U}_2]'(n,n)
     -2id(n)
   \right\} \nonumber\\*
 &\qquad
    -\frac{1}{2}\xi(n) [\,\mathcal{U}_1, \mathcal{U}_2]'(n,n+\hat1+\hat2)
\Bigr] \nonumber\\*
 &= \kappa \sum_n \tr\Bigl[
     \frac{1}{2}
     \lambda(n) \bigl(Q \lambda(n)\bigr)^\dagger
   + \frac{1}{2}\xi(n) \bigl(Q\xi(n)\bigr)^\dagger
     \Bigr] . 
 \label{eq:un-lambda}
\end{align}
Because of the nilpotency, the invariance under 
the $Q$-transformation is manifest.
Note that from the last expression the bosonic part of the
action is positive (semi-)definite.
The overall factor $\kappa$ is given as\footnote{
Here we use a normalization $\tr(T^a T^b)=\delta^{ab}$
for the generators of gauge group.
 }
\begin{equation}
 \kappa=\frac{1}{g^2 a^2} = \frac{N}{\lambda a^2},
\end{equation}
where $g$ is a dimensionful gauge coupling, $\lambda=g^2/N$ is a
't~Hooft coupling, and $a$ is the lattice spacing.
Eq. (\ref{eq:un-action}) reproduces a continuum action with fermion
in a twisted basis:
\begin{align}
 S_{\rm cont.}
 &=\frac{1}{g^2}\int d^2x \tr\Bigl[
    \frac{1}{4}F_{\mu\nu}^2 + \frac{1}{2}\sum_{i=1,2} (D_\mu s_i)^2
     -\frac{1}{2}[s_1, s_2]^2 + \frac{1}{2}d^2 \nonumber\\
 &\qquad    -\lambda D_1 \alpha -\lambda D_2 \beta 
     +\xi D_1 \beta -\xi D_2 \alpha
 + \lambda [s_1,\alpha] + \lambda[s_2,\beta]
 + \xi[s_1, \beta] -\xi [s_2, \alpha]
\Bigr],
\end{align}
where we have expanded the bosonic link variables as
\begin{align}
 \mathcal{U}_\mu &= 1 + iaA_\mu + a s_\mu +\dots
\end{align}
and rescaled the dimensionless lattice fields $d \to a^2 d$
and $(\alpha,\beta,\lambda,\xi) 
\to (a^\frac{3}{2}\alpha,a^\frac{3}{2}\beta, 
a^\frac{3}{2}\lambda, a^\frac{3}{2}\xi)$. 
The covariant derivative is $D_\mu=\partial_\mu+i[A_\mu,\ \cdot]$
and the curvature is $F_{\mu\nu}=-i[D_\mu, D_\nu]$.
The continuum $Q$-transformation is the following:
\begin{align}
 Q A_1 & = -i\alpha,  \qquad Q \alpha = 0, \label{eq:Qalpha-cont}\\
 Q A_2 & = -i\beta,   \qquad Q \beta = 0, \label{eq:Qbeta-cont}\\
 Q s_1 &= \alpha,     \qquad\qquad Q s_2 = \beta, \\
 Q \lambda &= D_1 s_1 + D_2 s_2 -id, \\
 Qd &= -i[D_1,\alpha] -i[D_2,\beta] + i[s_1,\alpha] + i[s_2,\beta], \\
 Q\xi &= iF_{12} + [s_1,s_2] -D_1 s_2 + D_2 s_1.
 \label{eq:Qxi-cont}
\end{align}
The continuum action and $Q$-transformation are valid for both $U(N)$
and $SU(N)$ gauge group.

\section{Reduction to $SU(N)$}
\label{sec:sun}

A naive reduction of the gauge group from $U(N)$ to $SU(N)$  
does not work because of the following argument.
We assume that fermions are algebra valued.
Consider a fermion $\alpha(n)$, which should be traceless in the $SU(N)$
case.  
However, since its gauge transformation is
\begin{equation}
 \alpha(n) \rightarrow G(n)\alpha(n)G(n+\hat1)^{-1},
\end{equation}
it is no longer in general traceless after the transformation.
For the bosonic link fields it is not the case if we identify 
it using the exponential function\footnote{
We use the same $\mathcal{U}_\mu$ as in the $U(N)$ case but all
$\mathcal{U}_\mu$ hereafter are for the $SU(N)$.
}
\begin{equation}
 \mathcal{U}_\mu =\exp(aiA_\mu + as_\mu), \label{eq:exp-link}
\end{equation}
where $A_\mu$ and $s_\mu$ are the traceless Hermitian gauge field and
the scalar, respectively.\footnote{
A different type of exponential parameterization
$\mathcal{U_\mu} = H_\mu U_\mu$ was proposed in \cite{Unsal:2005yh}, 
where $H_\mu$ is a positive hermitian matrix and $U_\mu$ is a unitary
matrix.  This is equivalent to parameterize 
$\mathcal{U}_\mu =e^{as_\mu} e^{iaA_\mu}$.
}
Because of the traceless exponent the determinant
of $\mathcal{U}_\mu$ is unity.
Since the determinant of the gauge transformation $G(n)$ is unity, 
it does not change the determinant of $\mathcal{U}_\mu$ 
(see the gauge transformation (\ref{eq:link-gauge-trans}) ). 
In other words, a natural
expansion for the bosonic link fields for $SU(N)$ 
is not a linear one but the exponential.
Note that in the leading order in $a$, (\ref{eq:exp-link}) agrees with
the linear parameterization $\mathcal{U}_\mu = 1 + iaA_\mu + as_\mu$ but
cannot keep the traceless property of $A_\mu$ and $s_\mu$ 
under the gauge transformation. \footnote{
See \cite{Galvez:2012sv} for a comparison of the linear and exponential
parameterization in the $U(N)$ system.
}

\begin{figure}
 \hfil\includegraphics{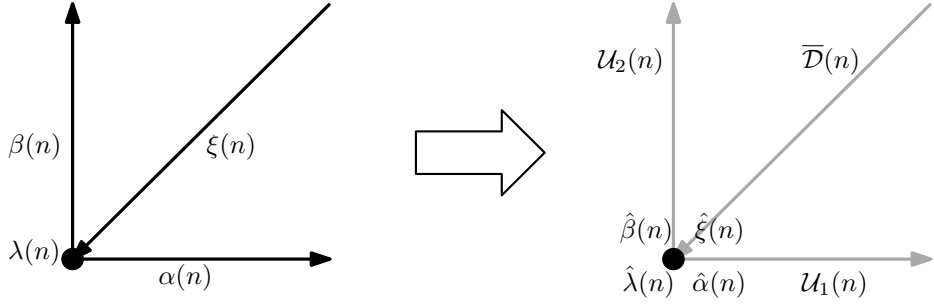}
 \label{fig:reducedfield}
 \caption{The link fermions (left) are decomposed into site fermions and
 link bosons (right).}
\end{figure}

Therefore we define fermions  on sites
in order to keep the traceless nature.
We decompose the link fermion into a product
of a site fermion and a link boson (see Fig.~\ref{fig:reducedfield}).
First, we decompose $\alpha(n)$ and $\beta(n)$ into
\begin{align}
 \alpha(n)&=\hat{\alpha}(n)\mathcal{U}_1(n), 
 & \beta(n)&=\hat{\beta}(n)\mathcal{U}_2(n),
\end{align}
where the hat ($\hat{\ }$) fields are defined on the site and 
can be expanded in the $SU(N)$ algebra:
\begin{align}
 \hat{\alpha}(n)&=\sum_a T^a \hat{\alpha}^a(n),
& \hat{\beta}(n)&=\sum_a T^a \hat{\beta}^a(n).
\end{align}
Their gauge transformation is
\begin{align}
 \hat{\alpha}(n) &\rightarrow G(n)\hat{\alpha}(n)G^{-1}(n),
 &
 \hat{\beta}(n) &\rightarrow G(n)\hat{\beta}(n)G^{-1}(n),
\end{align}
which is consistent with the transformation for $\alpha$, $\beta$
and $\mathcal{U}_\mu$.
From the $Q$-transformations for $\mathcal{U}_\mu$, $\alpha$ and $\beta$,
we obtain 
the $Q$-transformations for the hatted fields:
\begin{align}
 Q\hat{\alpha}(n) &= 2\hat{\alpha}(n)\hat{\alpha}(n),
 &
 Q\hat{\beta}(n) &= 2\hat{\beta}(n)\hat{\beta}(n).
 \label{eq:Qfermions}
\end{align}
After the $Q$-transformation they are still traceless, 
because of the fermionic nature,
\begin{align}
 \hat{\alpha}(n)\hat{\alpha}(n)
 &= \sum_{a,b} \hat{\alpha}^a(n)\hat{\alpha}^b(n) T^a T^b 
  = \sum_{a,b} \hat{\alpha}^a(n)\hat{\alpha}^b(n) \frac{1}{2}[T^a,T^b],
\end{align}
and the commutator of $SU(N)$ generators $[T^a, T^b]$ being traceless.
The nilpotency $Q^2\hat\alpha=Q^2\hat\beta=0$ follows from
the fermionic nature of $Q$: 
$Q^2\hat{\alpha} 
= 2(Q\hat{\alpha}) \hat{\alpha} -2 \hat{\alpha}(Q\hat{\alpha}) =0$, 
for example.

It should be noted that the $Q$-transformations in eq.~(\ref{eq:Qfermions})
are higher order terms in lattice spacing: they are order $a$ terms.
Although they are non-linear form at finite lattice spacing,
they give the same transformation as eqs.(\ref{eq:Qalpha-cont}) 
and (\ref{eq:Qbeta-cont}) in the continuum limit.  
This is the same feature as $Q$-transformation for the Sugino model.

We also introduce $\hat{\lambda}$,
a traceless part of the scalar fermion $\lambda$.
In order to keep it traceless, the $Q$-transformation should be
\begin{equation}
 Q\hat{\lambda}(n)
 =-\frac{1}{2}\left\{ 
    [\Ubar_1, \mathcal{U}_1]'(n,n) + [\Ubar_2,\mathcal{U}_2]'(n,n)
    \right\}_{\rm t.l.}
    -i\hat{d}(n),
\end{equation}
where we have also introduced a traceless auxiliary field $\hat{d}$
and $\rm t.l.$ refers to a traceless part
\begin{equation}
 \{A\}_{\rm t.l.} \equiv A-\frac{1}{N}\tr(A).
\end{equation}
Note that the shifted commutator $[\, \cdot\, ,\,\cdot\, ]'$ is not traceless.
The gauge transformation for $\hat{\lambda}$ and $\hat{d}$ is a one for
site variables:
\begin{align}
 \hat{\lambda}(n) &\rightarrow G(n)\hat{\lambda}(n)G^{-1}(n),
&  \hat{d}(n) &\rightarrow G(n)\hat{d}(n)G^{-1}(n).
\end{align}
The $Q$-transformation of the auxiliary field $\hat{d}$ 
is obtained by requiring 
the nilpotency for $Q^2 \hat\lambda=0$:
\begin{equation}
 Q\hat{d}(n) = i\left\{
          [\Ubar_1, \hat{\alpha} \mathcal{U}_1]'(n,n) 
          +[\Ubar_2, \hat{\beta} \mathcal{U}_2]'(n,n) 
	\right\}_{\rm t.l.}\, .
\end{equation}
Here we used the same transformation for $\Ubar_\mu$ as in the $U(N)$
case, $Q\Ubar_\mu = 0$.
We can show $Q^2 \hat{d}=0$ as well.

The treatment of the diagonal-link fermion $\xi$ is non-trivial,
because there is no diagonal-link boson.
We define the traceless site fermion $\hat{\xi}$ and a diagonal link boson
$\Dbar$ as
\begin{equation}
 \xi(n)=\Dbar(n)\hat{\xi}(n).
\end{equation}
Here we assume that $\Dbar$ is a function of $\Ubar_\mu$ thus $Q\Dbar=0$.
The gauge transformation for these fields is
\begin{align}
 \Dbar(n) &\rightarrow G(n+\hat1+\hat2)\Dbar(n)G(n)^{-1},
 &
 \hat{\xi}(n) &\rightarrow G(n)\hat{\xi}(n) G(n)^{-1}.
\end{align}
From eq.~(\ref{eq:Qxi}) we define
the traceless $Q$ transformation of $\hat{\xi}$ as
\begin{align}
 Q\hat{\xi}(n)
  &=\left\{ 
    \Dbar(n)^{-1}[\Ubar_1, \Ubar_2]'(n+\hat1+\hat2,n) \right\}_{\rm t.l.} 
     \nonumber\\
  &= -i \left\{ \Dbar(n)^{-1} \Fbar_{12}(n) \right\}_{\rm t.l.},
   \label{eq:Qxihat}
\end{align}
where we have introduced
\begin{align}
 i\mathcal{F}_{12}(n)&=[\mathcal{U}_1, \mathcal{U}_2]'(n, n+\hat1+\hat2),
 &
 i\Fbar_{12}(n)&=[\Ubar_1, \Ubar_2]'(n+\hat1+\hat2,n).
\end{align}
The $Q$-transformations are summarized in Appendix~\ref{app:action}.

In obtaining the action, we keep its $Q$-exact structure.
It seems that it is just a rewriting of the preaction (\ref{eq:un-lambda})
in terms of site fermions.  As we will see later, 
this is not a case with $\xi$, however.

The first term in (\ref{eq:un-lambda}) becomes
\begin{align}
 \Lambda^{(1)}_{SU(N)}
 &\equiv \kappa \sum_n \tr\Bigl[
     \frac{1}{2}
     \hat{\lambda}(n) \bigl(Q \hat{\lambda}(n)\bigr)^\dagger\Bigr]
\end{align}
and the contribution to the bosonic part of the action is
\begin{align}
 S^{(1)}_{SU(N)}\biggr|_{\rm Bosonic}
 &= Q \Lambda^{(1)}_{SU(N)}\biggr|_{\rm Bosonic} \nonumber \\
 &= \kappa \sum_n \Bigl[
    \frac{1}{8}\tr\bigl|\left\{
      [\Ubar_1, \mathcal{U}_1]'(n,n) + [\Ubar_2, \mathcal{U}_2]'(n,n)
\right\}_{\rm t.l.}
     \bigr|^2 
    + \frac{1}{2}\tr \hat{d}(n)^2
    \Bigr].
 \label{eq:su-bosonic1}
\end{align}
One solution which gives the minimum of this term is unitary
$\mathcal{U}_\mu$ and $\Ubar_\mu$ ($=\mathcal{U}_\mu^\dagger$), 
With the parameterization in eq.~(\ref{eq:exp-link}),
the scalar fields $s_\mu$ are vanishing in this solution. 
Setting gauge fields $A_\mu$ zero and scalars constant, we also find another
solutions, which is consistent with the fact that this term is a part of
kinetic term of the scalar fields.

The second term of the preaction $\Lambda$ requires some care.
The first candidate of a $SU(N)$ version is just a replacement
of $\xi$ with $\Dbar\hat{\xi}$ in eq.~(\ref{eq:un-lambda}):
\begin{equation}
 \Lambda'^{(2)}_{SU(N)}
 = \kappa \sum_n \tr\Bigl[
    \frac{1}{2}\mathcal{D}(n)\Dbar(n)
     \hat{\xi}(n) \bigl(Q\hat{\xi}(n)\bigr)^\dagger
     \Bigr],
\end{equation}
where $\mathcal{D}=\Dbar^\dagger$ and we have used $Q\Dbar=0$.
The bosonic action from this term is
\begin{align}
  S'^{(2)}_{SU(N)}\biggr|_{\rm Bosonic}
 &= Q \Lambda'^{(2)}_{SU(N)}\biggr|_{\rm Bosonic} \nonumber\\
 &= \kappa \sum_n \frac{1}{2}\tr\left[
     \mathcal{D}(n)\Dbar(n)
      \left\{\Dbar(n)^{-1}\Fbar_{12}(n)\right\}_{\rm t.l.}
\left\{\mathcal{F}_{12}(n)\mathcal{D}(n)^{-1}\right\}_{\rm t.l.}
   \right] .
 \label{eq:su-S2-wrong}
\end{align}
$\mathcal{F}_{12}\mathcal{D}^{-1}$ should be a function of the plaquette
made of $\mathcal{U}_\mu$ and $\mathcal{U}_\mu^{-1}$ (but does not
contain $\Ubar_\mu$ or $\Ubar_\mu^{-1}$).  For simplicity, let us set
the scalar fields $s_\mu$ to zero which gives a unitary plaquette.
If the value of the plaquette belongs to a center of the gauge group, 
$\exp(\frac{2\pi i}{N}k)$ ($k=0,1,\dots, N-1$), its traceless part is
always zero
and gives a minimum of the action.  
Each of the plaquettes can have an arbitrary
value in the center so the vacuum is highly degenerated.
This is exactly the same situation found in a very first version of
Sugino model~\cite{Sugino:2003yb}.
The other part of the bosonic action (\ref{eq:su-bosonic1}) does not
help to resolve this degeneracy.

In order to suppress the extra vacua,
according to~\cite{Sugino:2004qd}, we make use of an admissibility condition.
We replace $\mathcal{D}\Dbar$ in the preaction
(\ref{eq:su-S2-wrong})
with a (divergent) function, 
and propose the following modification:
\begin{align}
 \Lambda^{(2)}_{SU(N)}
 &= \kappa \sum_n \tr\Bigl[
    \frac{1}{2}
    \frac{1}{1-\frac{1}{\epsilon^2}\frac{1}{N}||1-\mathcal{U}_{12}(n)||^2}
     \hat{\xi}(n) \bigl(Q\hat{\xi}(n)\bigr)^\dagger
     \Bigr],
\end{align}
where $\epsilon$ is a (small) number discussed in the next section,
\begin{equation}
 \mathcal{U}_{12}(n)
  \equiv 
  \mathcal{U}_1(n) \mathcal{U}_2(n+\hat1) 
  \mathcal{U}_1(n+\hat2)^{-1}\,\mathcal{U}_2(n)^{-1}
\end{equation}
and the squared norm of a matrix is
\begin{equation}
 ||A||^2 \equiv \tr (A^\dagger A).
\end{equation}
Now the new action from $\Lambda^{(2)}_{SU(N)}$ becomes
\begin{align}
  S^{(2)}_{SU(N)}\biggr|_{\rm Bosonic}
 &= Q \Lambda^{(2)}_{SU(N)}\biggr|_{\rm Bosonic} \nonumber\\
 &= \kappa \sum_n \frac{1}{2}\tr\left[
    \frac{1}{1-\frac{1}{\epsilon^2}\frac{1}{N}||1-\mathcal{U}_{12}(n)||^2}
  \left|
    \left\{\mathcal{F}_{12}(n)\mathcal{D}(n)^{-1}\right\}_{\rm t.l.}
  \right|^2
   \right] .
 \label{eq:su-bosonic2}
\end{align}
Here we have written only the bosonic part 
and omitted the fermionic part.
It is worth mentioning that 
neither $\mathcal{D}$ nor $\Dbar$ is needed any more
in the action and the $Q$-transformation.
What we need is only $\mathcal{D}^{-1}$ (and $\Dbar^{\,-1}$)
and we assume that
$-i\{\Dbar^{\,-1}(n)\Fbar_{12}(n)\}_{\rm t.l.}$
should go to the r.h.s of (\ref{eq:Qxi-cont})
in a continuum limit, because it is $Q\hat{\xi}$.

We further impose the admissibility condition so that the action is
\begin{align}
 S&= \begin{cases}
      Q\left(\Lambda^{(1)}_{SU(N)} + \Lambda^{(2)}_{SU(N)}\right) \qquad&
      \frac{1}{N}|| 1-\mathcal{U}_{12}(n)||^2 < \epsilon^2,
       \\
      \infty & \text{otherwise}. 
     \end{cases}
 \label{eq:final-action}
\end{align}
Note that $||1-\mathcal{U}_{12} || = \mathcal{O}(a^4)$ in a naive power
counting.\footnote{
This counting is too naive because $||1-\mathcal{U}_{12}||^2$ is
proportional to a divergent operator.  
The actual scaling should be $\mathcal{O}(a^2)$.  See the analysis in
the next section.
}
Therefore in a numerical simulation we expect that the admissibility 
condition is practically  always satisfied if the simulation is close enough to
the continuum limit.
The explicit action with fermionic part after $Q$-transformation is given 
in Appendix~\ref{app:action}.

A possible variation of the action is to include
$1/(1-\frac{1}{\epsilon^2} ||1-\mathcal{U}_{12}  ||^2)$ 
in $\Lambda^{(1)}_{SU(N)}$ as well. This makes the action more complicated
but symmetric: terms from both 
$Q\Lambda^{(1)}_{SU(N)}$ and $Q\Lambda^{(2)}_{SU(N)}$ are
 equally complicated.

The measure for the path integral is invariant under the
$Q$-transformation as well.  A similar argument to \cite{Sugino:2006uf}
shows that
\begin{equation}
 \left(\prod_\mu d[\mathcal{U}_\mu(n)]\right)
  d[\hat{d}(n)]\, 
  d[\hat\alpha(n)]\, d[\hat\beta(n)]\, d[\hat\lambda(n)]\, d[\hat\xi(n)]
\end{equation}
is the invariant measure.
Here, 
$d[({\rm field})(n)]= \prod_n \prod_{a=1}^{N^2-1} d({\rm field})^a(n)$
for the auxiliary field $\hat{d}$ and fermions $\hat\alpha, \hat\beta,
\hat\lambda, \hat\xi$.
The measure for complex gauge field
$d[\mathcal{U}_\mu(n)] = \prod_n d\mathcal{U}_\mu(n)$
requires some care.
We use the following parameterization (we suppress the suffix $\mu$ and
lattice coordinate $n$)
\begin{equation}
 \mathcal{U} = \exp(i\sum_{A=1}^{2N^2-2} X^A T^A)
  \qquad (X^A:\text{real},\ \tr(T^A)=0).
\end{equation}
Based on a similar argument to the standard Haar measure,
we define
\begin{equation}
 d\mathcal{U}=c \sqrt{ \det(g_{AB} + g_{BA})} \prod_A dX^A,
\label{eq:measure-for-U}
\end{equation}
where 
\begin{equation}
 g_{AB}
 =\tr\left[
   \mathcal{U}^{-1}\frac{\partial \mathcal{U}}{\partial X^A} 
   \left(\mathcal{U}^{-1} 
  \frac{\partial \mathcal{U}}{\partial X^B} \right)^\dagger\right]
\end{equation}
and $c$ is a normalization constant.
One can show that this measure is invariant under left-multiplication 
by a similar quantity, 
$\mathcal{V} = \exp(i\sum_{A=1}^{2N^2-2} Y^A T^A)$.
$Q$-transformation of $\mathcal{U}_\mu$ is this type of multiplication,
\begin{align}
 \mathcal{U}_1 
 &\to \mathcal{U}_1 + i\varepsilon Q \mathcal{U}_1 
   = \exp(2 i\varepsilon\hat{\alpha}) \mathcal{U}_1, &
 \mathcal{U}_2 
 &\to \mathcal{U}_2 + i\varepsilon Q \mathcal{U}_2 
   = \exp(2 i\varepsilon\hat{\beta}) \mathcal{U}_2
\end{align}
with a grassmann parameter $\varepsilon$,
so the measure is invariant under $Q$-transformation.
In order to keep the invariance of (\ref{eq:measure-for-U})
with the right-multiplication of $\mathcal{V}$,
we need $\mathcal{V}\mathcal{V}^\dagger =1 $.
That is, only multiplication of a unitary matrix keeps 
the right-invariance of the measure.
The gauge transformation, which is both right- and left- multiplications
of unitary matrix, keeps the measure invariant.
Note that in Monte Carlo simulations, 
updating of $\mathcal{U}_\mu$
is just a (left-) multiplying $\mathcal{V}$ 
as well so we do not need any measure term.

\section{Detailed analysis on the Degenerate Vacua}
\label{sec:vacua}

In this section we give a bound for $\epsilon$ which appears in the
admissibility condition.
We need more information on $\mathcal{D}^{-1}$ so we restrict ourselves to
the following case,
\begin{equation}
  \mathcal{D}(n)^{-1}
  = t\, \mathcal{U}_1(n+\hat2)^{-1}\,\mathcal{U}_2(n)^{-1}
    + (1-t)\, \mathcal{U}_2(n+\hat1)^{-1}\,\mathcal{U}_1(n)^{-1},
\end{equation}
where $0\leq t\leq 1$.  

First we look for solutions of
$\bigl\{i\mathcal{F}_{12}(n) \mathcal{D}(n)^{-1}\bigr\}_{\rm t.l.}=0$,
which minimize (\ref{eq:su-bosonic2}).
With our choice of $\mathcal{D}^{-1}$, we have
\begin{equation}
 i\mathcal{F}_{12}(n) \mathcal{D}(n)^{-1}
  = 1 -2t 
    + t\,\mathcal{U}_{12}(n) 
    - (1-t) \bigl(\mathcal{U}_{12}(n)\bigr)^{-1}.
 \label{eq:F12Dinv}
\end{equation}
Then from the calculations given in Appendix~\ref{app:eom},
we obtain the following solutions:
\begin{itemize}
 \item $t=0$ or $t=1$: the center of $SU(N)$
       \begin{equation}
	\mathcal{U}_{12}=
	\begin{pmatrix}
	 e^{\frac{2\pi i}{N}n}&\\
         && \ddots \\
	 &&& e^{\frac{2\pi i}{N}n}
	\end{pmatrix}
	\qquad (n:\text{integer})
	\label{eq:sol-center}
       \end{equation}
 \item $t\neq 0$ and $t\neq 1$:
       \begin{equation}
	\mathcal{U}_{12}=
	\begin{pmatrix}
	  e^{i\alpha +\beta} 1_{k\times k} \\
          & -\frac{1-t}{t}e^{-i\alpha -\beta} 1_{(N-k)\times(N-k)}
	\end{pmatrix}
	\label{eq:sol-general}
       \end{equation}
        with an integer $k\neq \frac{N}{2}$
       and
       \begin{align}
	\alpha &= \frac{N-k-2n}{N-2k}\pi \qquad (n:\text{integer}), &
         e^\beta =\left(\frac{1-t}{t}\right)^{\frac{N-k}{N-2k}}.
	\label{eq:sol-general2}
       \end{align}
       Setting $k=N$, we obtain the center of the $SU(N)$, which is unitary.
       As a special case $t=\frac{1}{2}$, we also obtain $e^{\beta}=1$ thus
       $\mathcal{U}_{12}$ is unitary.  
       \footnote{
       Setting $t=\frac{1}{2}$ and $\mathcal{U}_{12}$ to unitary,
       we obtain the same equation of motion 
       as Sugino model~\cite{Sugino:2004qd}.
       The solutions found in \cite{Sugino:2004qd} covers only
       $k=2l=2n$ case, which gives $\alpha=\pi$.
       A careful analysis of the Sugino model,
       however, proves that it
       has more solutions and they coincide with ours \cite{sugino}.
       The new solutions do
       not affect to the parameter for the admissibility condition
       obtained in~\cite{Sugino:2004qd}.
       }

 \item $t=\frac{1}{2}$ and $N=4m$:
       in addition to the above,
       \begin{equation}
	\mathcal{U}_{12}=
	\begin{pmatrix}
	 e^{i\alpha +\beta} 1_{2m\times 2m} \\
         &  -e^{-i\alpha -\beta} 1_{2m\times 2m}
	\end{pmatrix}
	\label{eq:sol-special}
       \end{equation}
	with arbitrary real parameters $\alpha$ and $\beta$.
       
\end{itemize}

Then we look for the maximum allowed value for $\epsilon^2$.
We use the admissibility condition
\begin{equation}
 \frac{1}{N}||1-\mathcal{U}_{12}(n)||^2 < \epsilon^2
 \label{eq:admissibility}
\end{equation}
for all the plaquette, and all solutions listed above
need to violate this condition except for $\mathcal{U}_{12}(n)=1$ solution.
Once one fixed $N$ and $t$, it is a straightforward task to
find the maximum value of $\epsilon^2$ but is not easy to give
a general form.  Here we consider only $t=0, 1$ and $\frac{1}{2}$ cases.

\begin{itemize}
 \item $t=0$ or $t=1$.
       Substituting the center element (\ref{eq:sol-center}) to 
       the l.h.s of eq.(\ref{eq:admissibility}), we obtain
       \begin{equation}
	\frac{1}{N}||1-\mathcal{U}_{12}(n)||^2
          =4 \sin^2 \frac{n\pi}{N}.
       \end{equation}
       Because $n=0$ is the one we want to keep, i.e., $\mathcal{U}_{12}=1$,
       in order to suppress the unwanted vacua we should
       chose
       \begin{equation}
	\epsilon^2 \leq 4\sin^2\frac{\pi}{N}.
       \end{equation}
 \item $t=\frac{1}{2}$.
       The solution (\ref{eq:sol-general}) gives
       \begin{equation}
	\frac{1}{N}||1-\mathcal{U}_{12}(n)||^2
	=\frac{4(N-2k)}{N}\sin^2\frac{k-2n}{2(N-2k)}\pi + \frac{4k}{N}
       \end{equation}
       and a special case $N=4m$ (\ref{eq:sol-special}) gives
       \begin{equation}
       \frac{1}{N}||1-\mathcal{U}_{12}(n)||^2
	=
	2\left(\sinh \beta - \frac{1}{2}\cos\alpha\right)^2
           +2 -\frac{1}{2}\cos^2\alpha
	   \geq \frac{3}{2}. \label{eq:sol-special2}
       \end{equation}
       Noting a symmetry under $k \to N-k$ in (\ref{eq:sol-general}),
       we obtain
       \begin{align}
	N&=2 &  \epsilon^2&\leq 4, &&&& ( k=0, n=1) \\
        N&=3 &  \epsilon^2&\leq \frac{8}{3}, &&&& ( k=1 ) \\
        N&=4 &  \epsilon^2&\leq \frac{3}{2}, &&&& ( \text{from eq.(\ref{eq:sol-special2})})\\
	N&\geq 5 & \epsilon^2&\leq 4\sin^2\frac{\pi}{N}, &&&& (k=0, n=1)
       \end{align}
       where inside the parentheses indicates from which solution the bound for
       $\epsilon^2$ comes.
\end{itemize}
For large $N$, the bound for $\epsilon^2$ scales 
as $1/N^2$.  This implies we need a finer
lattice for larger $N$.  We can estimate the scaling as follow.
In a naive continuum limit, the action becomes
\begin{equation}
 S\sim \frac{Na^2}{\lambda}\sum_n \sum_a (F_{12}^a(n))^2 + \cdots,
\end{equation}
where $F_{12}$ is a dimensionful gauge curvature and sum over 
color and coordinate indices are explicitly written.
This gives the scaling
\begin{equation}
 (F_{12}^a(n))^2\sim \frac{\lambda}{Na^2}
\end{equation}
and thus
\begin{equation}
 \frac{1}{N}|| 1- \mathcal{U}_{12}(n)||^2
  \sim \frac{a^4}{N}\sum_{a=1}^{N^2-1} (F_{12}^a(n))^2 + \cdots  \sim a^2 \lambda.
\end{equation}
Therefore the maximum lattice spacing should scale as
\begin{equation}
 a^2\lambda \sim \frac{1}{N^2}
\end{equation}
as $N$ becomes larger to satisfy the admissibility condition.

\section{Conclusions and Discussions}
\label{sec:conclusions}

We proposed a lattice action for $SU(N)$ $\mathcal{N}=(2,2)$ super
Yang-Mills (\ref{eq:final-action}), 
starting with the CKKU model which has link fermions.  
Because fermions defined on the link cannot be $su(N)$
algebra valued, we decompose link fermions into site fermions and link bosons.
We kept a nilpotent fermionic $Q$-transformation for these field 
at finite lattice spacing, which is a scalar part in terms of the
topological twist.
Using a fermionic nilpotent $Q$-transformation, 
we defined a $Q$-exact action.  
By construction the action enjoys $Q$-invariance manifestly.
We encountered the degenerate vacua of the lattice
model which does not capture the correct continuum physics.
To suppress the artifact vacua, 
we used an admissibility condition.
In Table~\ref{tab:difference2} we summarize
the feature of the model.
\begin{table}
\hfil
 \begin{tabular}{c|c|c|c}
  & Sugino& CKKU & $SU(N)$-CKKU\\
 \hline
 twist & A-model & B-model & B-model \\
 fermion & site & link & site\\
 gauge group & $U(N)$ or $SU(N)$ & $U(N)$ only  & $SU(N)$\\
 admissibility & needed & no need & needed
 \end{tabular}
\caption{Comparisons of models.}
\label{tab:difference2}
\end{table}

A potential difficulty of CKKU model is a stability of $U(1)$ part
of the scalar. 
In this paper, we removed the $U(1)$ part so there is no problem about
the stability.
However, due to the admissibility condition we introduced,
the action became rather complicated.
Removing the $U(1)$ part removed the problem with 
the stability of the $U(1)$ part of the scalar.  
We still need to take care of flat directions of the
potential for the $SU(N)$ part of the scalar, by using large enough $N$
or a mass term, for example.

It is interesting to introduce matter multiplet to this model.
A lattice model for 2-dimensional $\mathcal{N}=(2,2)$ supersymmetric
QCD in~\cite{Kadoh:2009yf} has similar degenerate vacua in
the pure gauge sector, but they are resolved thanks to effects by the
matter sector.  The same thing might apply to our model.

\subsection*{Acknowledgements}
The author thanks H.~Suzuki and D.~Kadoh for critical comments and
discussions on the preliminary work.  Discussions with F.~Sugino
were quite useful to complete this work.
He also thanks F.~Bruckmann, S.~Catterall and M.~Hanada. 
He thanks all the member of Theoretical Physics Laboratory in RIKEN
and Okayama Institute for Quantum Physics for their hospitalities
during his stay.
Discussions at YITP Workshop ``Field Theory and String Theory'' (YITP-W-11-05)
were useful to complete this work.
The author is supported in part by the DFG SFB/Transregio 55
and acknowledges support from the EU ITN STRONGnet.

\appendix
\section{Lattice Action}
\label{app:action}

In this appendix, we give the explicit expression of the 
action (\ref{eq:final-action}):
\begin{align}
 S&= \begin{cases}
      Q\left(\Lambda^{(1)}_{SU(N)} + \Lambda^{(2)}_{SU(N)}\right) \qquad&
      \frac{1}{N}|| 1-\mathcal{U}_{12}(n)||^2 < \epsilon^2,
       \\
      \infty & \text{otherwise}.
     \end{cases}
\end{align}
$Q$-transformation given in Sec.~\ref{sec:sun} is
\begin{align}
 Q \mathcal{U}_1(n) 
  &= 2\hat{\alpha}(n)\,\mathcal{U}_1(n),
  \qquad Q\hat{\alpha}(n)= 2\hat{\alpha}(n)^2, \\
 Q \mathcal{U}_2(n) 
  &= 2\hat{\beta}(n)\,\mathcal{U}_2(n),
  \qquad Q\hat{\beta}(n)= 2\hat{\beta}(n)^2, \\
 Q\Ubar_1(n)&= Q\Ubar_2(n)=0, \\
 Q\hat{\lambda}(n)
  &= -\frac{1}{2}\left\{ 
      \bigl[\Ubar_1, \mathcal{U}_1)\bigr]'(n,n)
      +\bigl[\Ubar_2, \mathcal{U}_2)\bigr]'(n,n)
\right\}_{\rm t.l.} - i\hat{d}(n), \\
 Q \hat{d}(n)
  &= i\left\{
      \bigl[\Ubar_1, \hat{\alpha}\mathcal{U}_1\bigr]'(n,n)
      +\bigl[\Ubar_2, \hat{\beta}\mathcal{U}_2\bigr]'(n,n)
    \right\}_{\rm t.l.}, \\
Q\hat{\xi}(n)
  &=-i\left\{\Dbar(n)^{-1}\Fbar_{12}(n)\right\}_{\rm t.l.}.
\end{align}
The bosonic terms from $Q \Lambda_{SU(N)}^{(1)}$ and 
$Q \Lambda_{SU(N)}^{(2)}$ are given by eqs.~(\ref{eq:su-bosonic1})
and (\ref{eq:su-bosonic2}), respectively:
\begin{align}
 S^{(1)}_{SU(N)}\biggr|_{\rm Bosonic}
 &= \kappa \sum_n \Bigl[
    \frac{1}{8}\tr\bigl|\left\{
      [\Ubar_1, \mathcal{U}_1]'(n,n) + [\Ubar_2, \mathcal{U}_2]'(n,n)
\right\}_{\rm t.l.}
     \bigr|^2 
    + \frac{1}{2}\tr \hat{d}(n)^2
    \Bigr], \\
  S^{(2)}_{SU(N)}\biggr|_{\rm Bosonic}
 &= \kappa \sum_n \frac{1}{2}\tr\left[
    \frac{1}{1-\frac{1}{\epsilon^2}\frac{1}{N}||1-\mathcal{U}_{12}(n)||^2}
  \left|
    \left\{\mathcal{F}_{12}(n)\mathcal{D}(n)^{-1}\right\}_{\rm t.l.}
  \right|^2
   \right] .
\end{align}
The fermionic actions are 
\begin{equation}
 S^{(1)}_{SU(N)}\biggr|_{\rm Fermionic}
 = \kappa \sum_n \Bigl[
    \hat{\lambda}(n)[\Ubar_1,\hat{\alpha}\mathcal{U}_1]'(n,n)
    +\hat{\lambda}(n)[\Ubar_2,\hat{\beta}\mathcal{U}_2]'(n,n)
\Bigr]
\end{equation}
and
\begin{equation}
 S^{(2)}_{SU(N)}\biggr|_{\rm Fermionic} 
 = S_{\rm F}^{(2a)} + S_{\rm F}^{(2b)},
\end{equation}
where
\begin{align}
 S_{\rm F}^{(2a)}
 &= \kappa \sum_n \biggl[
   -\frac{1}{2}
    \frac{1}{1-\frac{1}{\epsilon^2}\frac{1}{N} || 1-\mathcal{U}_{12}(n) ||^2}
    \tr\Bigl[
    \hat{\xi}(n) 
    Q \Bigl( i\mathcal{F}_{12}(n)\mathcal{D}^{-1}(n)\Bigr)
   \Bigr]
\biggr], \\
 S_{\rm F}^{(2b)}
 &= \kappa \sum_n \Biggl[
     \left(\frac{1}{1-\frac{1}{\epsilon^2}\frac{1}{N} 
       ||1-\mathcal{U}_{12}(n) ||^2} \right)^2
      \tr\Bigl[ \hat{\xi}(n) 
      \Bigl(i\mathcal{F}_{12}(n)\mathcal{D}^{-1}(n)\Bigr) \Bigr]
    \nonumber\\
 &\qquad
   \times \frac{1}{\epsilon^2}\frac{1}{N}
    \tr\Bigl[\Bigl( 
  \mathcal{U}_1(n)\hat{\beta}(n+\hat1) \mathcal{U}_1^{-1}(n)\mathcal{U}_{12}(n)
  -\mathcal{U}_{12}(n)\hat{\beta}(n)
    \nonumber\\
 &\qquad\qquad
  -\mathcal{U}_{12}(n)\mathcal{U}_2(n)\hat{\alpha}(n+\hat2)\mathcal{U}_2^{-1}(n)
  +\hat{\alpha}(n)\mathcal{U}_{12}(n)
\Bigr) \Bigl( 1- \Ubar_{12}(n) \Bigr) \Bigr]
\Biggr].
\end{align}
We need to specify $i\mathcal{F}_{12}(n)\mathcal{D}(n)^{-1}$ to calculate 
$Q$-transformation in $S_{\rm F}^{(2a)}$.
Setting $i\mathcal{F}_{12}(n) \mathcal{D}(n)^{-1}$ as eq.~(\ref{eq:F12Dinv}),
\begin{equation}
 i\mathcal{F}_{12}(n) \mathcal{D}(n)^{-1}
  = 1 -2t 
    + t\,\mathcal{U}_{12}(n) 
    - (1-t) \bigl(\mathcal{U}_{12}(n)\bigr)^{-1},
\end{equation}
we obtain
\begin{align}
 S_{\rm F}^{(2a)}
 &= \kappa \sum_n
     \frac{1}{1-\frac{1}{\epsilon^2}\frac{1}{N} || 1-\mathcal{U}_{12}(n) ||^2}
   \nonumber\\ 
 &\qquad
   \times \biggl\{
   t \tr \Bigl[\hat{\xi}(n) \Bigl(
      \mathcal{U}_{12}(n)\mathcal{U}_2(n)\hat{\alpha}(n+\hat2)\mathcal{U}_2^{-1}(n)
       -\hat{\alpha}(n)\mathcal{U}_{12}(n) \Bigr) \nonumber\\
  &\qquad\qquad
      - \hat{\xi}(n)\Bigl(
     \mathcal{U}_1(n) \hat{\beta}(n+\hat1) \mathcal{U}_1^{-1}(n)\mathcal{U}_{12}(n)
       -\mathcal{U}_{12}(n)\hat{\beta}(n)
     \Bigr)\Bigr]
   \nonumber \\
 &\qquad
  +(1-t) \tr \Bigl[
     \hat{\xi}(n) \Bigl(
   \mathcal{U}_2(n)\hat{\alpha}(n+\hat2) \mathcal{U}_2^{-1}(n)\mathcal{U}_{12}^{-1}(n)
    -\mathcal{U}_{12}^{-1}(n)\hat{\alpha}(n)
    \Bigr) \nonumber\\
 &\qquad\qquad
  -\hat{\xi}(n)\Bigl(
   \mathcal{U}_{12}^{-1}(n)\mathcal{U}_1(n)
    \hat{\beta}(n+\hat1)\mathcal{U}_1^{-1}(n)
   -\hat{\beta}(n)\mathcal{U}_{12}^{-1}(n)
 \Bigr)
\Bigr]\biggr\}.
\end{align}
To obtain the above expressions, we dropped irrelevant trace less symbols
($\rm t.l.$)  because of a relation
$\tr\bigl[\hat{\xi}(A)_{\rm t.l.} \bigr] = \tr\bigl[\hat{\xi} A \bigr]$
for any $N\times N$ matrix $A$ and traceless $\hat\xi$.
\section{Solution of the equation of motion}
\label{app:eom}

We solve the following equation:
\begin{equation}
0= \left\{ i\mathcal{F}_{12}(n) \mathcal{D}(n)^{-1} \right\}_{\rm t.l.}
  = \left\{
    t\,\mathcal{U}_{12}(n) 
    - (1-t) \bigl(\mathcal{U}_{12}(n)\bigr)^{-1}
    \right\}_{\rm t.l.} \ .
  \label{eq:eom}
\end{equation}
We parameterize $\mathcal{U}_{12}$ as
\begin{equation}
 \mathcal{U}_{12}(n)
 = \begin{pmatrix}
	e^{i\alpha_1 +\beta_1}  &&&  0\\
        &  e^{i\alpha_2 +\beta_2} \\
        &&\ddots \\
        0 &&& e^{i\alpha_N +\beta_N}
   \end{pmatrix},
 \label{eq:diagonal}
\end{equation}
with real parameters $\alpha_i, \beta_i$ which
satisfy $\sum_i \alpha_i = \sum_i \beta_i =0$.
Denoting the diagonal component as
\begin{align}
 \lambda_i 
  &= t e^{i \alpha_i + \beta} + (t-1) e^{-i\alpha_i-\beta_i} \nonumber\\*
  &= (t e^{\beta_i} - (1-t)e^{-\beta_i})\cos \alpha_i
       + i (te^{\beta_i} +(1-t)e^{-\beta_i})\sin \alpha_i
  \equiv s_i \cos\alpha_i + i c_i \sin \alpha_i,
\end{align}
we rewrite the solution of eq.~(\ref{eq:eom}) as
\begin{equation}
 \lambda_1=\lambda_2=\dots=\lambda_N 
\label{eq:propto1_a}
\end{equation}
which gives
\begin{align}
 \begin{cases}
  & s_i \cos\alpha_i = s_j \cos\alpha_j\\
  & c_i \sin\alpha_i = c_j \sin\alpha_j
 \end{cases} \label{eq:app}
\end{align}
for $i,j=1,\dots,N$.  Using $c_i^2 -s_i^2 = 4t(1-t)$ and 
$c_i \leq 2\sqrt{t(1-t)} \sin\alpha_i$, we obtain
\begin{equation}
 c_i \pm 2 \sqrt{t(1-t)}\sin \alpha_i
  =c_j \pm 2 \sqrt{t(1-t)}\sin \alpha_j.
\end{equation}
This gives
\begin{align}
 \begin{cases}
  & c_i = c_j \\
  & \sin\alpha_i = \sin\alpha_j,
 \end{cases}
\end{align}
which is valid even with $t=0$ or $1$.
 (Note that $c_i \neq 0$ and eq.~(\ref{eq:app}).)
Consistent solutions with eq.~(\ref{eq:app}) are
\begin{equation}
 \alpha_i=\alpha_j, \qquad \beta_i = \beta_j
\end{equation}
or
\begin{equation}
 \alpha_i = \pi -\alpha_j, \qquad e^{\beta_i}=\frac{1-t}{t}e^{-\beta_j}.
  \qquad (t\neq 0, 1)
\end{equation}
Suppose we set $\alpha_i$ to
\begin{equation}
 \alpha_i = ( \underbrace{\alpha,\alpha,\dots,\alpha}_{k}, \
              \underbrace{\pi-\alpha,\dots \pi-\alpha}_{N-k})
\end{equation}
and thus
\begin{equation}
 \beta_i = ( \underbrace{\beta,\beta,\dots,\beta}_{k},\
            \underbrace{-\beta +\ln\frac{1-t}{t},\dots,
	                -\beta +\ln\frac{1-t}{t}}_{N-k}),
\end{equation}
where $0\leq k \leq N$.
The additional condition from $SU(N)$ gives
\begin{equation}
 0 = \sum_{i=1}^N \beta_i
   = -(N-2k)\beta + (N-k) \ln\frac{1-t}{t},
\end{equation}
and
\begin{equation}
 2\pi m
   = \sum_{i=1}^N \alpha_i
   = -(N-2k)\alpha  + (N-k)\pi,
\end{equation}
where $m$ is an integer.
We finally obtain
\begin{align}
 N&\neq 2k :& &\alpha= \frac{N-k-2m}{N-2k}\pi,\quad
               \beta=\frac{N-k}{N-2k}\ln\frac{1-t}{t}, \\
 N&=2k:     & &\alpha, \beta \text{ are arbitrary},\ 
              t=\frac{1}{2},\ N=4m,
\end{align}
which give eq.~(\ref{eq:sol-general})--(\ref{eq:sol-special}).

If $t=0$ or $1$, only $k=N$ is possible. 
The solution becomes the center of $SU(N)$:
\begin{equation}
 \alpha_i=\frac{2\pi}{N}n, \quad \beta_i = 0.
\end{equation}

\end{document}